\documentclass[11pt]{article}
\usepackage{amssymb}
\usepackage{amsmath}
\usepackage{color}
\usepackage{verbatim}
\usepackage[colorlinks,linkcolor=blue,citecolor=green]{hyperref}
\newenvironment{proof}[1][Proof]{\textbf{#1.} }{\ \rule{0.5em}{0.5em}}

\usepackage{graphicx}
\allowdisplaybreaks[3]

\title{Differential constraints for hyperbolic systems \\ through $k-$Riemann invariants}
\author{Alessandra Jannelli$^1$, Natale Manganaro$^1$,  Alessandra Rizzo$^2$\\
\\
 \small  $^1$ Department of Mathematical, Computer,\\
  \small Physical and Earth Sciences (MIFT)\\
  \small University of Messina, \\
  \small V.le F. Stagno D'Alcontres 31, 98166 Messina, Italy \\
\small e.mail: ajannelli@unime.it (A. Jannelli), nmanganaro@unime.it (N. Manganaro)\\
 \small  $^2$ Department of Mathematics and Computer Science,\\
  \small University of Palermo, \\
\small via Archirafi 34, 98123 Palermo, Italy \\
\small e.mail: alessandra.rizzo07@unipa.it}
\date{}

\begin{document}
\maketitle
\begin{abstract}
In this paper we develop a reduction procedure for determining exact wave solutions of first order quasilinear hyperbolic  one-dimensional nonhomogeneous systems. The approach is formulated within the theoretical framework of the method of differential constraints and it makes use of the $k-$Riemann invariants. The solutions obtained permit to characterize rarefaction waves also for nonhomogeneous models so that Riemann problems can be solved. Applications to the Euler system describing an ideal fluid with a source term are given.
\end{abstract}

\vspace{0.3cm}

\vspace{0.3cm}

\section{Introduction}
Over the years many mathematical methods have been proposed for determining exact solutions of partial differential equations (PDEs). Among others, the Method of Differential Constraints \cite{jan} permits to characterize wave solutions for hyperbolic systems. The main idea of the method is to look for exact solutions of  the original PDEs system which also  satisfy a further set of differential relations {\it (differential constraints)}. The resulting overdetermined system must be considered and its differential compatibility must be studied (for more details see \cite{mel1}). Without any further hypotesis, such a procedure is difficult, in general, to accomplish but it can be slightly simplified by requiring the  involutiveness of the resulting overdetermined system \cite{fsy}-\cite{rsy} (i.e. by requiring that by differentiation it is not possible to obtain any equations independent from the original PDEs). In the case of hyperbolic systems, the involutiveness condition gives solutions which, in principle, can be useful for solving nonlinear wave problems \cite{ms1}-\cite{mrv}. Furthermore, an interesting application of the method to a parabolic model was given in \cite{rive}. For further convenience we sketch briefly the method in the case of a one-dimensional hyperbolic system.  

We consider the  quasilinear system
\begin{equation}
\mathbf{U}_t + A\left( \mathbf{U} \right) \mathbf{U}_x=\mathbf{B}\left( \mathbf{U} \right) \label{hs}
\end{equation}
where $\mathbf{U} \in \mathbb{R}^N$ is the field vector, $A$ the $N \times N$ matrix coefficients, $\mathbf{B}  \in \mathbb{R}^N$ the source vector, while $t$ and $x$ denote, respectively, time and space coordinates. We assume the hyperbolicity (in the $t-$direction) of (\ref{hs}) and we denote with $\lambda^i \left( \mathbf{U} \right)$ the eigenvalues of $A$ (characteristic speeds) while the corresponding right and left eigenvectors are indicated, respectively, by $\mathbf{d}^i \left( \mathbf{U} \right)$ and $\mathbf{l}^i \left( \mathbf{U} \right)$. Moreover we assume $\lambda^i \neq \lambda^j, \; \; \forall i\neq j$ (i.e. system (\ref{hs}) is stricly hyperbolic). We choose $\mathbf{d}^i$ and $\mathbf{l}^i$ such that the orthonomal condition is satisfied $(\mathbf{d}^i \cdot \mathbf{l}^j= \delta^{ij})$.  It can be proved \cite{mel1} that the most general first order differential constraints which can be appended to (\ref{hs}) take the form
\begin{equation}
\mathbf{l}^\alpha \cdot \mathbf{U}_x=q^\alpha \left(\mathbf{U} \right) \label{vin1}
\end{equation}
where here and in the following we consider the case relevant for wave problems in which $\alpha=1,..,N-1$. By decomposing the vector $\mathbf{U}_x$ along the basis of the right eigenvectors $\mathbf{d}^i$ and taking (\ref{vin1}) into account, we have
\begin{equation}
\mathbf{U}_x=q^\alpha \mathbf{d}^\alpha+ \pi \mathbf{d}^N \label{ux}
\end{equation}
and, in turn, from (\ref{hs}) we find
\begin{equation}
\mathbf{U}_t=\mathbf{B}-\lambda^\alpha q^\alpha \mathbf{d}^\alpha - \lambda^N \pi \mathbf{d}^N \label{ut}
\end{equation}
where $\pi$ depends on $x$ and $t$. In (\ref{ux}), (\ref{ut}) and in the following we adopt the Einstein convention for which the sum for the repeated indices is implied. The involutiveness of the system (\ref{hs}), (\ref{vin1}) is assured if we require that the condition $\mathbf{U}_{tx}=\mathbf{U}_{xt}$ holds $\forall \pi$. In such a way we obtain $2(N-1)$ cumbersome compatibility equations for the $N-1$ unknown functions $q^\alpha (\mathbf{U})$. 

Finally, from (\ref{ux}) and (\ref{ut}) we obtain
\begin{equation}
\mathbf{U}_t+\lambda^N \mathbf{U}_x=\mathbf{B}+\left( \lambda^N - \lambda^\alpha \right) q^\alpha \mathbf{d}^\alpha. \label{equ}
\end{equation}
Since the left hand side of equations (\ref{equ}) involves the derivative of the field $\mathbf{U}$ along the characteristic curves associated to $\lambda^N$, system (\ref{equ}) can be integrated, in principle,  by means of the method of the characteristics. Therefore, once $q^\alpha$ have been determined from the compatibility conditions, by assigning the initial condition $\mathbf{U}(x,0)=\mathbf{U}_0 (x)$, integration of (\ref{equ}) gives the exact solution of (\ref{hs}) we are looking for, while it can be proved (see for instance \cite{mel1}) that the constraints (\ref{vin1}) specialize to
\begin{equation}
\mathbf{l}^\alpha \left( \mathbf{U}_0 \left( x \right) \right) \cdot \frac{d\mathbf{U}_0}{dx}=q^\alpha \left( \mathbf{U}_0 (x) \right). \label{vin2}
\end{equation}
Since the initial conditions $\mathbf{U}(x,0)=\mathbf{U}_0 (x)$ are constrained by the $N-1$ conditions (\ref{vin2}), the solution obtained from (\ref{equ}) is given in terms of one orbitrary function. It could be of a certain interest to notice that in the case where $q^\alpha =0$ and $\mathbf{B}=0$ the solutions obtained by (\ref{equ}) and (\ref{vin2}) are the well known simple waves.

Therefore, a key point of such a method is to study the compatibility conditions between (\ref{hs}) and (\ref{vin1}) whose solution is, in general, a hard task to accomplish. Within such a framework, the main contribution of this paper is to give an alternative strategy that simplifies the analysis of the compatibility conditions arising from  (\ref{hs}) and (\ref{vin1}). In fact, under suitable hypotheses, we are able to give the solutions of such  compatibility relations and, in turn, exact solutions of (\ref{hs}) are determined. Such a class of solutions can be useful for  characterizing rarefaction waves also for non-homogeneous systems like (\ref{hs}). This helps in studying Riemann problems and generalized Riemann problems also for systems of balance laws. We will apply  the procedure here considered for determing wave solutions for the hyperbolic model describing an ideal fluid with a source force term for which a Riemann problem is solved in terms of rarefaction waves. Some final remarks are given concerning the asymptotic behaviour of the rarefaction waves here characterized.

 The paper includes five sections. In Section 2 we illustrate our procedure and we show how the use of the $k-$Riemann invariants is helpful for studying the compatibility conditions and finding exact wave solutions for the governing nonhomogeneous hyperbolic system. In Section 3 we apply such an approach to the Euler equation describing an ideal fluid with a source term. In Section 4 we characterize rarefaction waves for the Euler system which permit to solve Riemann problems and some remarks are given about a conjecture proposed by Ruggeri et al. \cite{ruggeri1}, \cite{ruggeri2}. Some final comments are given in Section 5.

\section{Differential constraints through $k-$Riemann invariants}
Here our aim is to reformulate the procedure related to the Method of Differential Constraints 
in order to simplify it. Let us consider one of the characteristic speeds of (\ref{hs}) (for instance we fix, without loss of generality, $\lambda^N$) and we compute its Riemann invariants $R\left( \mathbf{U} \right)$ defined by
\begin{equation}
\mathbf{\nabla}R \cdot \mathbf{d}^N =0 \label{ri11}
\end{equation}
where $\nabla=\frac{\partial}{\partial \mathbf{U}}$. It is well known that associated to $\lambda^N$ there exist $N-1$ Riemann invariants $R^\alpha$ whose gradients are linearly independent (see for instance \cite{smo}). Therefore, owing to (\ref{ri11}), we can write
\begin{equation}
\nabla R^\alpha = \sigma_{\beta}^{\alpha}\,  \mathbf{l}^\beta, \quad \alpha, \beta = 1,..., N-1 \label{nr}
\end{equation}
where $\sigma_{\beta}^{\alpha}$ are the components of $\nabla R^\alpha$ with respect the basis of the left eigenvectors. Moreover, here and what follows, the greek indices vary from $1$ to $N-1$. Taking (\ref{nr}) into account, the constraints (\ref{vin1}) assume the form
\begin{equation}
\frac{\partial R^\alpha}{\partial x}=\sigma_{\beta}^{\alpha} \, q^\beta \label{rvin1}
\end{equation}
while from (\ref{equ}) we have
\begin{equation}
\frac{\partial R^\alpha}{\partial t}+ \lambda^N \frac{\partial R^\alpha}{\partial x}= \sigma_{\beta}^{\alpha} \, \mathbf{l}^\beta \cdot \mathbf{B}+ \left( \lambda^N - \lambda^\beta \right)\sigma_{\beta}^{\alpha} \, q^\beta \label{requ}
\end{equation}
Next, after choosing one of the field variables of $\mathbf{U}$ (say for instance $v=u_j$), we add to the $N-1$ equations (\ref{requ}) the $j-$th equation arising from (\ref{equ})
\begin{equation}
\frac{\partial v}{\partial t}+ \lambda^N \frac{\partial v}{\partial x}=B_j + \left( \lambda^N - \lambda^\beta \right)q^\beta d_{j}^{\beta} \label{vequ}
\end{equation}
where we indicate with $B_j$ and $d_{j}^{\alpha}$ the $j-$th component, respectively, of $\mathbf{B}$ and $\mathbf{d}^\alpha$. Thus, under the variable transformation
\begin{equation}
R^\alpha=R^\alpha (\mathbf{U}), \quad v=u_j \label{transf}
\end{equation}
the equations (\ref{equ}) transform to (\ref{requ}) and (\ref{vequ}) while the constraints (\ref{vin1}) take the form (\ref{rvin1}). Integration of (\ref{requ}), (\ref{vequ}) along with (\ref{rvin1}) gives, through the change of variables (\ref{transf}), exact solutions of (\ref{hs}), (\ref{vin1}). To this end we require the involutiveness of the overdetermined system (\ref{rvin1})-(\ref{vequ}) (i. e.  we require that the condition $\partial_{tx}R^\alpha=\partial_{xt} R^\alpha$ holds $\forall  \, \partial_x v$). The resulting compatibility conditions are:
\begin{eqnarray}
&&\left( \lambda^\beta - \lambda^N \right) \sigma_{\beta}^{\alpha} \, \frac{\partial q^\beta }{\partial v}=\left( \left( \lambda^N - \lambda^\beta \right) \frac{\partial \sigma_{\beta}^{\alpha}}{\partial v} - \sigma_{\beta}^{\alpha} \frac{\partial  \lambda^\beta }{\partial v} \right) q^\beta +\frac{\partial}{\partial v} \left( \sigma_{\beta}^{\alpha} \, \mathbf{l}^\beta \cdot \mathbf{B} \right) \label{c1} \\
&& \frac{\partial w^\alpha}{\partial R^\gamma} \, z^\gamma - \frac{\partial z^\alpha}{\partial R^\gamma} \, w^\gamma +\frac{\partial w^\alpha}{\partial v}\left( B_j + \left( \lambda^N - \lambda ^\gamma \right) q^\gamma d_{j}^{\gamma}\right)=0 \label{c2}
\end{eqnarray}
where we set
\begin{equation}
w^\alpha= \sigma_{\beta}^{\alpha} \, q^\beta, \quad \quad z^\alpha=\sigma_{\beta}^{\alpha} \, \mathbf{l}^\beta \cdot \mathbf{B}- \lambda^\beta \sigma_{\beta}^{\alpha} \, q^\beta. \label{wz}
\end{equation}
 The $N-1$ equations (\ref{c1}) characterize a linear ODE-like system in the unknown $q^\alpha$ which, due the strictly hyperbolicity of (\ref{hs}), can be written in normal form. Once the functions $q^\alpha$ are determined from (\ref{c1}), substituting them in (\ref{c2}), we find a set of $N-1$ structural conditions to which the coefficients of the system (\ref{hs}) must be satisfied to garantee the compatibility among (\ref{hs}) and (\ref{vin1}).

In the following some cases where (\ref{c1}) and (\ref{c2}) can be solved are presented.

\vspace{0.2cm}
\noindent
{\it i)} We assume $q^\alpha =0$ so that from (\ref{c1}), (\ref{c2}) we find
\begin{equation}
\sigma_{\beta}^{\alpha} \, \mathbf{l}^\beta \cdot \mathbf{B}= F^\alpha (R^\gamma) \label{co1}
\end{equation}
where $F^\alpha$ are not specified functions. If the structural condition (\ref{co1}) is satisfied, then, from (\ref{rvin1}) we find $R^\alpha=R^\alpha (t)$ and taking (\ref{requ}), (\ref{vequ}) into account,  exact solutions of (\ref{hs}), (\ref{vin1}) are obtained by solving the system
\begin{eqnarray}
&&\frac{d R^\alpha}{d t}=F^\alpha (R^\gamma) \label{c1eq1}, \quad \quad  \\
&&\frac{\partial v}{\partial t} + \lambda^N\left( v, R^\alpha (t) \right) \frac{\partial v}{\partial x}=B_j \left( v, R^\alpha (t) \right). \label{c1eq2}
\end{eqnarray}
In passing we notice that the equations (\ref{c1eq1}) are decoupled from (\ref{c1eq2}). In fact, once $R^\alpha (t)$ are determined from (\ref{c1eq1}), exact solutions of the governing system can be obtained by solving the quasilinear non-autonomous PDE (\ref{c1eq2}). Furthermore, when $\mathbf{B}=0$ also $F^\alpha =0$ and the equations (\ref{c1eq1}), (\ref{c1eq2}) characterize the simple waves
\begin{eqnarray}
&&R^\alpha \left( \mathbf{U} \right)=k^\alpha  \nonumber \\
&& u_j=v=v_0 (\xi), \quad \quad x=\lambda^N \left( v_0 \left( \xi \right), k^\alpha \right)t +\xi \nonumber
\end{eqnarray}
where $k^\alpha$ are arbitrary constants and $v_0 (x)=v(x,0)=u_j (x,0)$.

\vspace{0.2cm}
\noindent
{\it ii)} We now require
\begin{equation}
\frac{\partial \lambda^N}{\partial v}=0 \quad \Rightarrow \quad \lambda^N (R^\gamma). \label{la}
\end{equation}
In such a case from (\ref{c1}) we can calculate $q^\alpha$ by solving the algebraic equations
\begin{equation}
\left(\lambda^\beta - \lambda^N \right) \sigma_{\beta}^{\alpha} \, q^\alpha =\sigma_{\beta}^{\alpha} \, \mathbf{l}^\beta \cdot \mathbf{B}+ F^\alpha (R^\gamma) \label{cc1}
\end{equation}
where $F^\alpha$ are not specified functions. Next, by substituting  $q^\alpha$ in (\ref{c2}), where $z^\alpha$ assumes the form
$$
z^\alpha=-\left( F^\alpha +\lambda^N \sigma_{\beta}^{\alpha} \, q^\beta \right),
$$
we get the structural conditions which must be satisfied in order that the procedure considered in the present case holds. Thus, the equations (\ref{requ}) specializes to
\begin{equation}
\frac{\partial R^\alpha}{\partial t}+\lambda (R^\gamma) \frac{\partial R^\alpha}{\partial x}=-F^\alpha (R^\gamma) \label{rr}
\end{equation}
which can be integrated indipendently from (\ref{vequ}). Once $R^\alpha (x,t)$ are determined from (\ref{rr}), $v(x,t)$ can be obtained by integration of (\ref{vequ}) which assumes a semilinear form because of (\ref{la}).

\vspace{0.2cm}
\noindent
{\bf Remark 1.} It could be of a certain interest to notice that the condition (\ref{la})  gives $\nabla \lambda^N \cdot \mathbf{d}^N =0$ so that the characteristic speed $\lambda^N$ is exceptional \cite{boillat} (or linearly degenerate \cite{lax}) and the solution of (\ref{hs}) which can be obtained in case {\it ii)} exists for $ \forall t \geq 0$. Viceversa if we assume that $\lambda^N$ is linearly degenerate we find $\frac{\partial \lambda^N}{\partial v} d_{j}^{N}=0$, where $d_{j}^{N}$ indicates the $j-$th component of $\mathbf{d}^N$. Therefore, if $d_{j}^{N}  \neq 0$, we find condition (\ref{la}). 

\vspace{0.2cm}
\noindent
{\it iii)} We require
\begin{equation}
\sigma_{\beta}^{\alpha} \left( \mathbf{l}^\beta \cdot \mathbf{B} - \lambda^\beta q^\beta \right)= F^\alpha (R^\gamma), \quad \quad \sigma_{\beta}^{\alpha}\, q^\beta = G^\alpha (R^\gamma). \label{cc2}
\end{equation}
Owing to (\ref{cc2}), the condition (\ref{c1}) is identically satisfied, while from (\ref{c2}) we find
\begin{equation}
\frac{d G^\alpha}{d R^\beta}\,  F^\beta -\frac{d F^\alpha}{dR^\beta}\, G^\beta=0. \label{k1}
\end{equation}
Since $q^\alpha$ can be calculated from (\ref{cc2})$_2$, the relations (\ref{cc2})$_1$ and (\ref{k1}) are the structural conditions which must be satisfied by the coefficients of (\ref{hs}) for its compatibility with (\ref{vin1}). 

In the present case, equations (\ref{requ}) assumes the form
\begin{equation}
\frac{\partial R^\alpha}{\partial t}+ \lambda^N \frac{\partial R^\alpha}{\partial x}= F^\alpha +\lambda^N G^\alpha \label{e31}
\end{equation}
while the constraints (\ref{vequ}) specializes to
\begin{equation}
\frac{\partial R^\alpha}{\partial x}=G^\alpha. \label{e32}
\end{equation}
Of course, if $\lambda^N (R^\gamma)$ as in {\it ii)} case, system (\ref{e31}) is decoupled from (\ref{vequ}).

\section{Exact solutions for the Euler equations}
In this section we apply the approach illustrated in the previous one for determining exact solutions for the Euler equations describing an ideal fluid
\begin{eqnarray}  
&&\rho_t + u \rho_x + \rho u_x =0 \label{e1} \\
&&u_t + u u_x + \frac{c^2}{\rho} \rho_x + \frac{p_s}{\rho} S_x= f(\rho, u) \label{e2} \\
&& S_t + u S_x =0 \label{e3}
\end{eqnarray}
where $\rho$ is the mass density, $u$ the velocity, $S$ the entropy, $p(\rho, S)$ the pressure, $c = \sqrt{p_\rho}$ the sound velocity and $f(\rho, u)$ a force term. System (\ref{e1})-(\ref{e3}) is strictly hyperbolic, the characteristic speeds are $\lambda^1=u-c$, $\lambda^2=u$, $\lambda^3=u+c$, while the corresponding orthornormal left and right eigenvectors write as follows
\begin{eqnarray}
&&\mathbf{l}^1=\begin{pmatrix}
\frac{c^2}{\rho} & -c & \frac{p_s}{\rho}
\end{pmatrix}, \quad    \mathbf{l}^2=\begin{pmatrix}
0 & 0 & 1
\end{pmatrix}, \quad \mathbf{l}^3=\begin{pmatrix}
\frac{c^2}{\rho} & c & \frac{p_s}{\rho}
\end{pmatrix} \label{m1} \\
&& \mathbf{d}^1=\frac{1}{2c^2}\begin{pmatrix}
\rho & -c & 0
\end{pmatrix}^T, \quad    \mathbf{d}^2=-\frac{1}{c^2}\begin{pmatrix}
p_s & 0 & -c^2
\end{pmatrix}^T, \quad \mathbf{d}^3=\frac{1}{2c^2}\begin{pmatrix}
\rho & c & 0
\end{pmatrix}^T \label{m2}
\end{eqnarray}
In the following, first we point out our attention to the case $\lambda^N=\lambda^3$ and after we will choose $\lambda^N=\lambda^2$. The analysis of the case $\lambda^N=\lambda^1$ can be carried on as that corresponding to $\lambda^3$ and therefore we will not consider it. 

\subsection{Wave solutions associated to $\lambda^3=u+c$}
When $\lambda^N=\lambda^3=u+c$, the corresponding Riemann invariants are
\begin{equation}
R^1=u-\int{\frac{c}{\rho}d\rho}, \quad \quad R^2=S \label{ri1}
\end{equation}
while from (\ref{nr}) we have
$$
\sigma_{1}^{1}=-\frac{1}{c}, \quad \sigma_{2}^{1}=\frac{p_s}{\rho c}-\int{\frac{c_s}{\rho}d\rho}, \quad \sigma_{1}^{2}=0, \quad \sigma_{2}^{2}=1. 
$$ 
Next we choose $v=u$, so that equation (\ref{requ}) and (\ref{vequ}) assume the form
\begin{eqnarray}
&&\frac{\partial R^1}{\partial t}+ \lambda^3 \frac{\partial R^1}{\partial x}=f -2q^1 + \left(\frac{p_s}{\rho}-c\int{\frac{c_s}{\rho}d\rho} \right)q^2 \label{eu1} \\
&&\frac{\partial R^2}{\partial t}+\lambda^3 \frac{\partial R^2}{\partial x}= c \, q^2 \label{eu2} \\
&&\frac{\partial v}{\partial t}+\lambda^3 \frac{\partial v}{\partial x}=f-q^1 \label{eu3}
\end{eqnarray}
while the differential constraints (\ref{rvin1}) assume the form
\begin{equation}
\frac{\partial R^1}{\partial x}=-\frac{q^1}{c}+\frac{q^2}{c}\left(\frac{p_s}{\rho}-c \int{\frac{c_s}{\rho}\rho} \right), \quad \quad \frac{\partial R^2}{\partial x}=q^2 \label{eu4}
\end{equation}
where $q^1 (R^1, R^2, v)$ and $q^2 (R^1, R^2, v)$ are still unspecified.

In the present case conditions (\ref{c1}) specializes to:
\begin{eqnarray} 
&& 2\frac{\partial q^1}{\partial v}=\frac{\partial f}{\partial v}+\frac{q^1}{c}\frac{\partial (v+c)}{\partial v}+c q^2 \frac{\partial}{\partial v}\left( \frac{p_s}{\rho c} - \int{\frac{c_s}{\rho}d\rho}\right) \label{q2} \\
&&c \frac{\partial q^2}{\partial v}=q^2 \label{q1}
\end{eqnarray}
while relation (\ref{c2}) for $\alpha=2$ gives
\begin{equation}
\left( f - q^1 \right) \left( \frac{\partial q^2}{\partial R^1}+\frac{\partial q^2}{\partial v}\right) =0 \label{q3}
\end{equation}
From (\ref{q3}) we consider the case
\begin{equation}
q^1=f \label{q4}
\end{equation}
so that, from (\ref{q2}), (\ref{q1}) we have
\begin{eqnarray}
&&q^2= \rho F(R^1, R^2) \label{qq1} \\
&&f=\rho F(R^1, R^2) \left( \frac{p_s}{\rho}- c \int{\frac{c_s}{\rho}d\rho}\right)+ \rho \, c \, G(R^1, R^2) \label{qq2} 
\end{eqnarray}
where $F(R^1, R^2)$ and $G(R^1, R^2)$ are not specified. Furthermore it is simple to verify that condition (\ref{c2}) with $\alpha=1$ are identically satisfied by (\ref{q4}) and (\ref{qq1}), taking (\ref{qq2}) into account. Of course, once $p(\rho, S)$ and/or $f(\rho, u)$ are assigned, the structural condition (\ref{qq2}) must be checked to be consistent. To this end, here and in the following we point our attention to three celebrated constitutive pressure laws.
\begin{itemize}
\item The ideal gas pressure law \cite{cf}
\begin{equation}
p=A(S) \rho^\gamma, \quad \quad A(S)=e^{\frac{S-\hat{S}}{C_v}} \label{rg}
\end{equation}
where  $C_v$ is the specific heat at constant volume, $\gamma > 1$ is the gas index and $\hat{S}$ is a constant.
\item The Von-K\'arm\'an law \cite{vk}
\begin{equation}
p=-\frac{a^2 (S)}{\rho}+b(S) \label{vk}
\end{equation}
where $a(S)$ and $b(S)$ are smooth functions.
\item The Chaplygin gas pressure law \cite{kmp}
\begin{equation}
p=-\frac{a_0^2 }{\rho} \label{ck}
\end{equation}
where $a_0$ is constant.
\end{itemize}
 
For instance, we consider the case of a ideal gas. In such a case, in order to relation (\ref{qq2}) to be consistent, it follows
\begin{equation}
f=\frac{k_2}{1-\gamma}\rho^\gamma+ \left( k_0 u + k_1 \right) \rho^{\frac{\gamma +1}{2}}, \quad F=\frac{k_2-2k_0  \sqrt{ \gamma A}}{A^\prime(S)}, \quad G=\frac{k_0 R^1 + k_1}{\sqrt{\gamma A}} \label{ccc}
\end{equation}
where $k_0$, $k_1$ and $k_2$ are constants while the prime stays for ordinary differentiation. Furthermore integration of equations (\ref{eu1})-(\ref{eu4}) leads to the following cases.

\noindent
{\it I)} If $k_0=k_1=0$ we get
\begin{eqnarray}
&&\rho= \left( \frac{\gamma -1}{2\sqrt{\gamma A(S)}}\left( u_0 (\xi) - c_1 \right)  \right)^{\frac{2}{\gamma-1}}, \quad \quad  u=u_0 (\xi) \label{s1} \\
&&  \nonumber \\
&&A(S) =\left( c_2 \left( u_0 (\xi) -c_1 \right) t+ \left( A\left( S_0 \left(\xi \right) \right) \right)^{\frac{\gamma}{\gamma -1}}\right)^{\frac{\gamma -1}{\gamma}}  \label{s3}
\end{eqnarray}
where $c_1$ is a constant while we set
$$
c_2=\frac{k_2 \gamma \sqrt{\gamma}}{\gamma -1} \left( \frac{\gamma -1}{2 \sqrt{\gamma}} \right)^{\frac{\gamma +1}{\gamma -1}}.
$$
Moreover, taking (\ref{rvin1}) into account,  the initial data $\rho(x,0)=\rho_0(x)$, $u(x,0)=u_0 (x)$ and $S(x,0)=S_0(x)$ must satisfy the constraints
\begin{equation}
\frac{d}{dx} A\left( S_0 \left( x \right) \right)=k_2 \rho_0 (x), \quad \quad \rho_0 (x) = \left( \frac{\gamma -1}{2 \sqrt{\gamma A\left( S_0 \left( x \right) \right)}}\left( u_0 (x) -c_1\right)\right)^{\frac{2}{\gamma -1}}. \label{cccc}
\end{equation}
Furthermore the wave variable $\xi (x,t)$ is defined implicitely by
$$
x=\left( \frac{\gamma +1}{2} u_0 (\xi) -  \frac{\gamma -1}{2} c_1 \right) t  + \xi.
$$

\noindent
{\it II)} If $k_0=k_2=0$ we obtain
\begin{equation}
\rho= \frac{\rho_0(\xi)}{1-\frac{k_1}{\sqrt{\gamma A_0}}\rho_0(\xi)t}, \quad u=u_0 (\xi), \quad S=S_0 \label{eee} 
\end{equation}
where $S_0$ is a constant, $A_0 =A(S_0)$, while, from (\ref{rvin1}), the inital conditions $\rho(x,0)=\rho_0 (x)$ and $u(x,0)=u_0(x)$ are related by
$$
\frac{d u_0 (x)}{dx}-\sqrt{\gamma A_0} \left( \rho_0 (x) \right)^{\frac{\gamma -3}{2}}\frac{d\rho_0 (x)}{dx}=-\frac{k_1}{\sqrt{ \gamma A_0}}\rho_0 (x).
$$
Furthermore $\xi (x,t)$ is given by
\begin{eqnarray}
&&x=u_0(\xi) t-\frac{2 \sqrt{\gamma A_0} } { k_1 \left( 3 -\gamma \right)} \rho_{0}^{\frac{\gamma - 3}{2}}  \left\{ \left( 1- \frac{k_1 \rho_0(\xi)}{\sqrt{\gamma A_0}}t \right)^{\frac{3-\gamma}{2}}-1\right\}+\xi \quad \mbox{if} \quad \gamma \neq 3 \label{x1} \\
&& \nonumber \\
&& x= u_0 (\xi) t- \frac{ \sqrt{3A_0}}{k_1} \ln{\left( 1- \frac{k_1}{\sqrt{3 A_0}} \rho_0 (\xi) t \right)}+ \xi \quad \mbox{if} \quad \gamma=3. \label{x2}
\end{eqnarray}

\vspace{0.3cm}
\noindent
Now we consider the  pressure laws (\ref{vk}) which when $b=0$ and $a=constant.$ specializes to (\ref{ck}).

It is known that system (\ref{e1})-(\ref{e3}) supplemented by (\ref{vk}) or (\ref{ck}) is completely exceptional (i. e. all its characteristic velocities are linearly degenerate). In the case where $a^\prime (S) \neq 0$,  from (\ref{qq2}) we find
\begin{equation}
f=c_0 \, u-\frac{c_1}{\rho}. \quad \quad F=\frac{c_0 \, a(S) + c_1}{a(S) a^\prime (S)}, \quad \quad G= \frac{c_0 \, R^1}{a(S)}-\frac{\left( c_0 a(S)+c_1\right) }{a^2(S) a^\prime (S)} \, b^\prime (S) \label{aa}
\end{equation} 
where $c_0$ and $c_1$ are constants, while from (\ref{requ}), (\ref{vequ}) we obtain
\begin{eqnarray}
&&\rho=a(S) \left\{ e^{-c_0 t} \left( \int_{0}^{t}{H\left( S\left(\xi, t \right)\right)e^{c_0 t}dt}+R_{0}^{1}(\xi)  \right) -u_0 (\xi) \right\}^{-1}, \quad \quad u=u_0 (\xi)  \label{j1} \\
&& a(S)=e^{c_0 t} \left\{ -\frac{c_1}{c_0} \left( e^{-c_0 t} -1 \right) + a_0 ( \xi )    \right\} \quad \mbox{if} \; c_0 \neq 0 \label{j2} \\
&&a(S)=c_1 t + a_0 (\xi) \quad \mbox{if} \; c_0 =0 \label{j3}
\end{eqnarray}
In (\ref{aa})-(\ref{j3}) we set
$$
H(S)=\frac{c_0 \, a +c_1}{a \, a^\prime} \, b^\prime (S), \quad \quad a_0 (\xi)=a\left( S_0 (\xi) \right), \quad \quad R_{0}^{1}(\xi)= u_0 (\xi)+\frac{a_0 (\xi)}{\rho_0 (\xi)}
$$
with $\rho(x,0)=\rho_0(x)$, $u(x,0)=u_0(x)$ and $S(x,0)=S_0(x)$. Furthermore $\xi(x,t)$ is given by
\begin{eqnarray}
&&x=-\frac{R_{0}^{1}(\xi)}{c_0} \left( e^{-c_0 t} -1 \right)+ \int_{0}^{t}{h(\xi,t) e^{-c_0 t}dt}+\xi \quad \mbox{if} \; c_0 \neq 0 \label{d1} \\
&&x=R_{0}^{1}(\xi)t+\int_{0}^{t}{h(\xi, t) dt}+\xi \quad \mbox{if} \; c_0=0 \label{d2}
\end{eqnarray}
where 
$$
h(\xi,t)=\int_{0}^{t}{H(S(\xi,t))e^{c_0t}}.
$$
Of course the initial data $\rho_0 (x)$, $u_0 (x)$ and $S_0(x)$ must satisfy the constraints (\ref{rvin1}).

Next we consider the case where $a=a_0=const.$ and $b=0$, so that condition (\ref{qq2}) gives
$$
f=\Psi (R^1), \quad \quad G=\frac{\Psi(R^1)}{a_0}
$$
while $F(R^1, R^2)$ is arbitrary. Once $\Psi(R^1)$ is assigned and $F(R^1, R^2)$ is choosed, we can integrate (\ref{eu1})-(\ref{eu3}) along with (\ref{eu4}). As an example, if $\Psi=c_0 \left( R^{1} \right)^2$ with $c_0=const.$ and $F=\frac{c_0}{a_0}R^1 R^2$, we find
\begin{eqnarray}
&&\rho=a_0 \left(  \frac{\rho_0(\xi)+c_0 \left( a_0+\rho_0(\xi) u_0(\xi)\right)t}{a_0-c_0 u_0(\xi) \left( a_0 + \rho_0(\xi) u_0(\xi) \right)t} \right), \quad u=u_0(\xi), \quad \label{h1} \\
&& \nonumber \\
&& S=S_0(\xi) \left\{ 1+c_0 t \left( u_0 (\xi) + \frac{a_0}{\rho_0 (\xi)} \right) \right\}\label{h2}
\end{eqnarray}
with
$$
x=\frac{1}{c_0}\ln{\left( 1+c_0 t \left( u_0 (\xi)+\frac{a_0}{\rho_0(\xi)}\right) \right)}+\xi
$$
while, owing to (\ref{rvin1}), the initial conditions $\rho(x,0)=\rho_0(x)$, $u(x,0)=u_0(x)$ and $S(x,0)=S_0(x)$ must obey to
\begin{equation}
\frac{d}{dx}\left( u_0 (x)+ \frac{a_0}{\rho_0 (x)}\right)=-c_0 \frac{\rho_0(x)}{a_0} \left( u_0(x)+ \frac{a_0}{\rho_0(x)} \right)^2, \quad \quad S_0(x)=\hat{c} \left( u_0(x)+ \frac{a_0}{\rho_0(x)}\right)^{-1} \label{i}
\end{equation}
with $\hat{c}=const.$

\vspace{0.3cm}
Now we assume $q^1 \neq f$. We are going to solve the compatibility conditions between (\ref{eu1})-(\ref{eu3}) and (\ref{eu4}) in the cases {\it i)-iii)} of the previous section. Let us start with the case {\it i)}. It follows soon that from condition (\ref{co1}) we obtain
\begin{equation}
f=F(R^1), \quad \quad p(\rho, S)=p_0 (\rho) + p_1 (S) \label{l}
\end{equation}
where $p_0(\rho)$ and $p_1(S)$ are unspecified functions, while from (\ref{requ}), (\ref{vequ}) and (\ref{rvin1}) we get
\begin{eqnarray}
&&\int{\frac{dR^1}{F(R^1)}\, dt}=t+t_0 \quad \Rightarrow \quad u-\int{\frac{c(\rho)}{\rho}d\rho}=\hat{R^1}(t) \label{m1} \\
&&u=\int_{0}^{t}{F \left( \hat{R}^1 (t) \right)}\, dt +u_0 (\xi), \quad \quad S=S_0=const. \label{m2}
\end{eqnarray}
where $t_0$ is a constant and $\xi(x,t)$ is given by
$$
x=u_0 (\xi) t+\int_{0}^{t}{\left( c \left( \rho\left( \xi, t \right) \right)+\int_{0}^{t}{F(\hat{R}^1 (t))\, dt}\right)  dt}+ \xi.
$$

\vspace{0.2cm}
\noindent
Now we consider the {\it ii)} case, so that from (\ref{la}), according to Remark 1, the pressure $p=p(\rho,S)$ specializes to the Von-K\'arm\'an law (\ref{vk}) or, as a special case, to (\ref{ck}). Therefore, excluding the case where $q^1 =f$ previously considered, from (\ref{c2}) and (\ref{cc1}) we have
\begin{eqnarray}
&&q^1=H(R^1)+ \left( \frac{a^\prime}{a}(R^1 -v)- \frac{b^\prime}{a}\right) F^2(R^2)+F^1(R^1, R^2) \label{a1} \\
&&q^2=\frac{F^2(R^2)}{R^1 -v}  \label{a2} \\
&&f=H( R^1)+q^1, \quad \quad \mbox{with} \quad H\neq 0 \label{a3}
\end{eqnarray}
where the functions $F^1(R^1, R^2)$, $F^2(R^2)$ and $H(R^1)$ are not specified.  Furthermore, equations (\ref{rr}) and (\ref{eu3}) assume the form
\begin{eqnarray}
&&\frac{\partial R^1}{\partial t} + R^1 \frac{\partial R^1}{\partial x }=-F^1(R^1, R^2) \label{g1} \\
&&\frac{\partial R^2}{\partial t} + R^1 \frac{\partial R^2}{\partial x }=-F^2(R^2) \label{g2} \\
&&\frac{\partial v}{\partial t} + R^1 \frac{\partial v}{\partial x }=H(R^1) \label{g3}
\end{eqnarray}
while the constraints (\ref{rvin1}) write
\begin{equation}
\frac{\partial R^1}{\partial x}=-\frac{H+F^1}{R^1 -v}, \quad \quad \frac{\partial R^2}{\partial x}=-\frac{F^2 }{R^1 -v }. \label{vvv}
\end{equation}
Since the force $f$ depends on $\rho$ and $u$, by checking the consistency of condition (\ref{a3}) we find the following cases
\begin{enumerate}
\item $a=a_0=const.$, $f(R^1)=2H(R^1)+g(R^1)$ and
$$
F^1=\frac{b^\prime(R^2)}{a_0}F^2(R^2)+g(R^1)
$$
where $g(R^1)$ is unspecified.
\item $a^\prime(S) \neq 0$ and
$$
F^2=-\frac{k_0 \, a(S) +k_1}{a^\prime (S)}, \quad F^1=\frac{b^\prime(S)}{a(S)}F^2-2H(R^1)+k_0 R^1, \quad f=k_0 u-\frac{k_1}{\rho}.
$$  
\end{enumerate}
where $k_0$ and $k_1$ are constants. Once $H(R^1)$, $g(R^1)$ and $F^2(R^2)$ are assigned in the first case or $H(R^1)$ in the second one, then integration of (\ref{g1})-(\ref{g3}) gives a class of exact solutions for the Euler equations supplemented, respectively, by the Chaplygin pressure law or by the Von-K\'arm\'an law while, the corresponding initial conditions must obey (\ref{vvv}).

\vspace{0.2cm}
\noindent
Next we consider the {\it iii)} case. From (\ref{cc2}), (\ref{k1}) we have
\begin{eqnarray}
&&F^2=G^2=0, \quad F^1=m(R^2) \, G^1 (R^1, R^2), \label{bb1} \\
&& q^1=-c \, G(R^1, R^2), \quad q^2=0 \label{bb2} \\
&&f=\left( u-c+m\left(R^2\right) \right) G^1(R^1, R^2) \label{bb3}
\end{eqnarray} 
where $m(R^2)$ and $G^1 (R^1, R^2)$ are unspecified. Of course we can check the consistency of (\ref{bb3}) once $p(\rho, S)$ is assigned. For instance, if we choose $p=A(S) \rho^\gamma$, from (\ref{bb3}) we find $f=const.$  with $\gamma=3$. Conversely, if we consider (\ref{vk}), condition (\ref{bb3}) gives
$$
a=a_0=const., \quad m=m_0=const., \quad \frac{\partial G^1}{\partial R^2}=0
$$
so that
$$
f=\left( u-\frac{a_0}{\rho}+m_0 \right) G^1 (R^1)
$$
and, by assuming $b=0$, the pressure law under concerns is (\ref{ck}).
The corresponding solution is
\begin{equation}
\frac{a_0}{\rho}=\left( \psi(\xi)+m_0 \right) \left(1 - \phi(z)\right), \quad  S=S_0=const, \quad u=\phi(z) \left( \psi(\xi) +m_0 \right)-m_0 \label{v1} 
 \end{equation}
where
$$
\xi=x+m_0 t, \quad \quad z=t- \int{\frac{d\xi}{\psi(\xi)+m_0}}.
$$
while, by assigning the initial data $\rho(x,0)=\rho_0 (x)$, $u(x,0)=u_0(x)$, the functions $\psi(\xi)$ and $\phi(z)$ must satisfy  the conditions
$$
\psi(x)=u_0(x)+ \frac{a_0}{\rho_0(x)}, \quad \phi \left( z_0 \left(x \right) \right)=\frac{u_0(x)+m_0}{\psi(x)+m_0} \quad \mbox{with}\quad z_0(x)=-\left( \int{\frac{d\xi}{\psi(\xi)+m_0}}\right)_{\xi=x}.
$$

\subsection{Wave solutions associated to $\lambda^2=u$}
The Riemann invariants corresponding to $\lambda^2=u$ are
\begin{equation} 
R^1=p(\rho, S), \quad \quad R^2=u \label{rr1}
\end{equation}
so that from (\ref{nr}) we find
$$
\sigma_{1}^{1}=\sigma_{3}^{1}=\frac{\rho}{2}, \quad \quad   \sigma_{1}^{2}=  -\sigma_{3}^{2}=-\frac{1}{2c}.
$$
In the following, without loss of generality, we choose $v=S$. Therefore, since $\lambda^2=u$ is linearly degenerate, we are lead to develop the analysis illustrated in case {\it ii)} of section 1. Thus, from (\ref{cc1}) we have
\begin{equation}
q^1=f+F^2(R^1, R^2)-\frac{F^1(R^1,R^2)}{\rho c}, \quad \quad q^3=f+F^2(R^1, R^2)+\frac{F^1 (R^1, R^2)}{\rho c}  \label{q1q2}
\end{equation}
while relations (\ref{c2}) assume the form
\begin{eqnarray}
&&F^1 \frac{\partial}{\partial R^1} \left( \rho \left( f+F^2 \right)\right)+F^2 \frac{\partial}{\partial R^2}\left( \rho \left( f+ F^2\right)\right)= \nonumber \\
&&=\rho \left(f + F^2 \right) \frac{\partial F^1}{\partial R^1}+\frac{F^1}{\rho c^2} \left( \frac{\partial F^1}{\partial R^2}  +\rho \left( f+ F^2 \right)\right) \label{com1} \\
&& \nonumber \\
&& F^1 \frac{\partial}{\partial R^1}\left( \frac{F^1}{\rho c^2} \right)+F^2 \frac{\partial}{\partial R^2} \left( \frac{F^1}{\rho c^2}\right)= \nonumber \\
&&=\rho \left( f+ F^2 \right) \frac{\partial F^2}{\partial R^1}+\frac{F^1}{\rho c^2} \left( \frac{\partial F^2}{\partial R^2} +\frac{F^1}{\rho c^2}\right) \label{com2}
\end{eqnarray}
and, in turn, equations (\ref{vequ}), (\ref{rr}) specializes to
\begin{eqnarray}
&&\frac{\partial R^1}{\partial t}+\lambda^2 \frac{\partial R^1}{\partial x}=-F^1 (R^1, R^2) \label{m1} \\
&&\frac{\partial R^2}{\partial t}+\lambda^2 \frac{\partial R^2}{\partial x}=-F^2 (R^1, R^2) \label{m2} \\
&& \frac{\partial v}{\partial t}+\lambda^2 \frac{\partial v}{\partial x}=0 \label{m3}
\end{eqnarray}
along with the constraints (\ref{rvin1})
\begin{equation}
\frac{\partial R^1}{\partial x}=\rho \left( f + F^2 \right), \quad \quad \frac{\partial R^2}{\partial x}=\frac{F^1}{\rho c^2}.  \label{vinc}
\end{equation}
Therefore, once $F^1$ and $F^2$ are determined from (\ref{com1}), (\ref{com2}), exact solutions of the Euler equations (\ref{e1})-(\ref{e3}) will be obtained by solving (\ref{m1})-(\ref{m3}) supplemented by (\ref{vinc}). To this end, in the following, we consider different cases.

\vspace{0.2cm}
\noindent
{\it I)} If $F^1=F^2=0$, conditions (\ref{com1}), (\ref{com2}) are satisfied and integration of (\ref{m1})-(\ref{m3}) gives
\begin{equation}
\rho=\rho_0(\xi), \quad u=u_0=const., \quad S=S_0(\xi) \quad \quad \mbox{with} \quad \xi=x-u_0 t  \label{tr}
\end{equation} 
where, owing to (\ref{vinc}), the initial data $\rho(x,0)=\rho_0(x)$, $S(x,0)=S_0(x)$ are related by the condition
$$
\frac{\partial}{\partial x}\left( p \left( \rho_0 (x), S_0 (x) \right) \right)=\rho_0 (x) f\left( \rho_0 (x), u_0 \right).
$$ 

\vspace{0.2cm}
\noindent
{\it II)} If $F^1=0$, by excluding the trivial case $f=-F^2$, we find $F^2=F^2(R^2)$ and
\begin{equation}
f(\rho, u)=-F^2 (R^2)+\frac{\psi (R^1 , v)}{\rho} \label{jj}
\end{equation}
where $\psi(R^1, v)$ is still unspecified. In the present case, from (\ref{m1})-(\ref{vinc}) we find
\begin{equation}
p\left( \rho, S \right)=p\left( \rho_0(\xi), S_0(\xi) \right), \quad \int{\frac{du}{F^2(u)}}=-t \; \Rightarrow \; u=\hat{u}(t), \quad S=S_0(\xi) \label{ss1}
\end{equation}
with 
$$
\xi=x-\int_{0}^{t}{\hat{u}(t) \, dt}
$$
while, the inital data $\rho(x,0)=\rho_0(x)$, $S(x,0)=S_0(x)$ are related by 
$$
\frac{d}{dx}\left( p \left( \rho_0(x), S_0(x) \right)\right) =\psi \left( p_0(x), S_0(x) \right).
$$
Of course,  the function $\psi (R^1, v)$ must be assigned such that the consistency of condition (\ref{jj}) is assured. For instance, from (\ref{jj}),  in the case of an ideal gas (\ref{rg}) we find $\psi=\psi(\rho^\gamma)$ while for  the Von-K\'arm\'an law (\ref{vk}) we deduce $\psi=\psi(\rho^{-1})$.

\vspace{0.2cm}
\noindent
{\it III)} We set $F^2=0$, so that from (\ref{com1}), (\ref{com2}) we find
\begin{equation}
F^1(p,u)=\rho^2 p_\rho \, \varphi (u, S), \quad \quad f(\rho, u)=\rho^2 p_\rho \left( \pi(u,S)-\frac{\varphi_u}{\rho} \right)  \label{x}
\end{equation}
where the functions $\varphi(u,S)$ and $\pi(u,S)$ are still unspecified. The conditions (\ref{x}) must be checked once $p(\rho, S)$ and/or $f(\rho,u)$ are assigned. For instance, if we consider  (\ref{rg}), we soon find $\varphi=A^\frac{1}{\gamma}$ and $\pi=\dfrac{\pi_0 (u)}{\gamma A}$ so that we get $f= \pi_0(u) \rho^{\gamma +1}$ and $F^1=\gamma p^{1+\frac{1}{\gamma}}$, where $\pi_0 (u)$ is a unspecified function. The resulting exact solution is
\begin{equation}
\rho=\frac{\rho_0(\xi) A_{0}^{-\frac{1}{\gamma}}}{\rho_0(\xi) t +A_{0}^{-\frac{1}{\gamma}}}, \quad u=u_0 (\xi), \quad S=S_0(\xi) \label{vv1} 
\end{equation}
where
$$
x=u_0(\xi)t + \xi, \quad \quad \mbox{and} \quad A_0=A\left(S_0\left( \xi \right) \right)
$$
while the initial conditions $\rho(x,0)=\rho_0(x)$, $u(x,0)=u_0(x)$ and $S(x,0)=S_0(x)$ must satisfy the constraints
$$
\frac{du_0(x)}{dx}=\left( A(S_0(x)) \right)^{\frac{1}{\gamma}}\rho_{0} (x), \quad \quad \frac{d}{dx}\left( A(S_0(x) \rho_{0}^{\gamma}(x)\right)= \pi_0 \left( u_0 \left( x \right)\right) \left( \rho_0 \left( x \right)\right)^{\gamma +2}.
$$

\vspace{0,2cm}
\noindent
{\it IV)} Now we assume $\rho \left( f+F^2 \right)=\mu(R^2)$, $F^2=F^2(R^2)$ and $F^1=F^1(R^1)$. In such a case condition (\ref{com2}) gives
\begin{equation}
F^2=k_0 R^2 +k_1, \quad \quad F^1=\rho c^2 \left( -k_0 + \rho h(S) \right) \label{vv3}
\end{equation}
where $k_0$, $k_1$ are constants while $\mu(R^2)$ and $h(S)$ are still unspecified. Next, we have to check the consistency of relation (\ref{vv3})$_2$ along with the remaining compatibility condition (\ref{com1}), first in the case of an ideal gas and after for the Von-K\'arm\'an pressure law.

If $p=A(S)\rho^\gamma$, it results $h(S)=k_1 A^{\frac{1}{\gamma} (S)}$ and the following two cases are found 
\begin{enumerate}
\item $k_1=0$, $F^1=-k_0 \gamma p$, $\mu=\mu_0 \left(R^2\right)^{-\left( \gamma +1 \right)}$, with $\mu_0=const.$ Consequently we have 
$$
f=-k_0 u + \mu_0 \frac{u^{-\left( \gamma + 1 \right)}}{\rho}.
$$
\item $\mu=0$, $F^1=\gamma \left( - k_0 p + k_1 p^{1+\frac{1}{\gamma}} \right)$ and, in turn, $f=-k_0 u-k_1$.
\end{enumerate}
In the first case we obtain the solution
\begin{equation}
\rho=\rho_0 (\xi) e^{k_0 t}, \quad u=-k_0 x, \quad S=\frac{\mu_0}{\gamma k_0} \left(  -k_0 \, \xi \, \rho_0 (\xi) \right)^{- \gamma} \quad \quad \mbox{with} \quad \xi=x \, e^{k_0 t}  \label{b3}
\end{equation}
while the initial data $\rho_0(x)$ is arbitrary.

In the second case, taking (\ref{rg}) into account, we find
\begin{eqnarray}
&&\rho=\frac{k_0 \rho_0 (\xi) e^{k_0 t}}{k_0 +k_1 \rho_0 (\xi) A_{0}^{\frac{1}{\gamma}}\left( e^{k_0 t} -1 \right)}, \quad u=-\frac{k_1}{k_0}+\left( u_0(\xi) +\frac{k_1}{k_0} \right) e^{-k_0 t}, \quad S=S_0(\xi) \label{k2} \\
&& \nonumber \\
&&\xi=\frac{k_0 \left( k_0 x+k_1 t \right)+k_1 \left( e^{-k_0 t} -1 \right)}{k_0 \left( k_0 + \hat{u}_0 \left( e^{- k_0 t}-1 \right)\right)} \label{k3}
\end{eqnarray}
with $A_0=A\left( S_0 \left( \xi \right) \right)$ while the initial conditions $\rho_0(x)$, $u_0(x)$ and $S_0(x)$ must obey to
\begin{equation}
u_0 (x)=\hat{u}_{0} \, x, \quad  S_0 (x)=-C_p \ln{\rho_0 (x)} \quad \quad \mbox{with} \quad \hat{u}_0=-k_0 +k_1 e^{-\frac{\hat{S}}{C_p}} \label{kj}
\end{equation}
where $C_p$ indicates the specific heat at constant pressure. Of course if $k_1=0$ and $\mu=0$ the above cases overlap.

\vspace{0.2cm}
If now we adopt the pressure law (\ref{vk}), from (\ref{vv3}) we obtain
$$
F^1(p)=k_0 p, \quad \quad h(S)=k_0 \frac{b(S)}{a^2 (S)}, \quad \quad \mu=0.
$$
Therefore $f=-(k_0 u +k_1)$ and the corresponding solution is
\begin{eqnarray}
&&\rho=\frac{a^2 \left( S_0 (\xi) \right)}{c_0 e^{-k_0 t}-b \left( S_0 (\xi) \right)}, \quad  u=-\frac{k_1}{k_0}+ \left( u_0 (\xi) + \frac{k_1}{k_0} \right) e^{-k_0 t}, \quad S=S_0(\xi) \label{p4} \\
&& \nonumber \\
&& x=-\frac{k_1}{k_0}t-\frac{1}{k_0}\left( u_0 (\xi) +\frac{k_1}{k_0} \right) \left( e^{-k_0 t}-1 \right) + \xi \label{p5}
\end{eqnarray}
with $c_0=const.$ Furthermore the initial data $\rho_0(x)$, $u_0(x)$ and $S_0(x)$ must satisfy the constraint
$$
\frac{du_0(x)}{dx}=k_0 \left( 1+\frac{b(S_0 (x) \rho_0 (x)}{a^{2}(S_0(x))} \right).
$$

\vspace{0.2cm}
\noindent
{\it V)} Finally we consider the case $\rho \left( f+F^2 \right)=\mu(R^2)$, $F^2=F^2 (R^2)$ and $F^1=F^1(R^2)$, with $\mu(R^2)$  unspecified function. From (\ref{com1}) and (\ref{com2}) we find $\mu=k_0$, $F^1=-k_0 u$, $F^2=\left( k_1 + k_2 u\right) F^1$, where $k_0$, $k_1$, $k_2$ are constants and, in turn, $f=k_0 \left(k_1 u + k_2 u^2 \right)+\frac{k_0}{\rho}$. Furthermore the following structural condition must be satisfied
\begin{equation}
\rho \frac{\partial p}{\partial \rho} \left(\rho H(S) -k_2 \right)=1 \label{gh}
\end{equation}
where $H(S)$ is arbitrary. After integration of (\ref{gh}), it results that, as far as we know, the only case physically acceptable is obtained when $k_2=0$. In fact, in such a case, from (\ref{gh}), the pressure  law (\ref{vk}) is found by setting $H(S)=\dfrac{1}{a^2 (S)}$. The resulting exact solution in terms of the field variables $(p, u, S)$ is
\begin{eqnarray}
&&p=\frac{u_0(\xi)}{k_1} \left( e^{k_0 k_1 t}- 1 \right) + p_0 (\xi), \quad u=u_0 (\xi) e^{k_0 k_1 t}, \quad S=S_0(\xi) \label{nm} \\
&& \nonumber \\
&&x=\frac{u_0(\xi)}{k_0 k_1} \left( e^{k_0 k_1 t}-1 \right) + \xi \label{nm1}
\end{eqnarray}
while the initial data $p\left( \rho_0 (x), S_0 (x) \right)=p_0 (x)$, $u_0(x)$ and $S_0(x)$ must satisfy the conditions
\begin{equation}
p_0(x)=k_0 x, \quad \quad \frac{du_0 (x)}{dx}=\frac{k_0 u_0(x)}{k_0 x -b\left( S_0(x) \right)}. \label{b3}
\end{equation}
Finally we notice that in the case of a Chaplygin gas, the solution given in (\ref{nm})-(\ref{b3}) specializes to
\begin{equation}
p=\frac{k_0 x \left( k_1 - \hat{c} \left( e^{k_0 k_1 t} -1 \right) \right)}{k_1 -\hat{c}\left( e^{k_0 k_1 t}-1 \right)}, \quad u=\frac{k_0 k_1 \hat{c}\,  x e^{k_0 k_1 t}}{\hat{c}\left( e^{k_0 k_1 t}-1 \right) -k_1}, \quad  S=S_0 (\xi) \label{fi2} 
\end{equation}
with
$$
\xi=\frac{k_1 x}{k_1 -\hat{c} \left( e^{k_0 k_1 t}-1 \right)}
$$
while the initial datum $S_0(x)$ is arbitrary.

\section{Nonlinear wave problems}
In this section our aim is to apply the reduction procedure developed in previous sections for studying  some relevant nonlinear wave problems.

\subsection{Riemann problem}
The famous Riemann problem (RP) is an initial data problem characterized by constant states with a discontinuity in a point  (say $x=0$). In the case of a hyperbolic system of conservation laws, the general solution of a RP was given by P. Lax \cite{lax}. He proved that, if the initial  states are not "far", then the solution of a RP is determined by constant states separated by rarefaction waves, shock waves and contact discontinuities. For nonhomogeneous systems such theory fails because, in general, rarefaction waves are not admitted by nonhomogeneous equations (balance laws). In fact rarefaction waves are characterized by simple wave solutions and such a class of exact solutions are usually admitted by homogeneous equations. Here, owing to the results previously obtained, we outline a possible approach for determining rarefaction waves also in the case of nonhomogeneous systems. 

Let us consider the Riemann problem
\begin{equation}
\mathbf{U}(x,0)=\left\{
	\begin{array}{l}
		\mathbf{U}_L \quad \quad \mbox{for} \; x<0   \\
		\\
		\mathbf{U}_R \quad \quad   \mbox{for} \; x>0
	\end{array}
	\right.  \label{ri1}
\end{equation}
where the constant states $\mathbf{U}_L \neq \mathbf{U}_R$ are equilibrium states of the governing system (\ref{hs}), so that
\begin{equation}
\mathbf{B}\left( \mathbf{U}_L \right)=\mathbf{B}\left( \mathbf{U}_R \right)=0. \label{eq}
\end{equation}
We require that $\mathbf{U}_L$ and $\mathbf{U}_R$ are also equilibrium states of the constraints (\ref{vin1}) and therefore we assume
\begin{equation}
q^\alpha \left( \mathbf{U}_L \right)=q^\alpha \left( \mathbf{U}_R \right) =0. \label{eq1}
\end{equation}
Taking the variable transformation (\ref{transf}) into account, from (\ref{eq}) and (\ref{eq1}) we have
$$
\mathbf{B}\left( R_{L}^{\gamma}, v_L \right)=\mathbf{B}\left( R_{R}^{\gamma}, v_R\right)=0, \quad \quad q^\alpha \left( R_{L}^{\gamma}, v_L\right)=q^\alpha \left( R_{R}^{\gamma}, v_R \right)=0
$$
where $R_{L}^{\alpha}=R^\alpha \left( \mathbf{U}_L \right)$, $R_{R}^{\alpha}=R^\alpha \left( \mathbf{U}_R \right)$. In the new variables $R^{\alpha}$ and $v$, the initial data (\ref{ri1}) transforms to
\begin{equation}
R^\alpha(x,0)=\left\{
	\begin{array}{l}
	 R_{L}^{\alpha}\quad  \mbox{for} \; x<0   \\
		\\
          R_{R}^{\alpha}  \quad   \mbox{for} \; x>0
	\end{array}
	\right.   \quad \quad 
v(x,0)=\left\{
          \begin{array}{l}
            v_L \quad \mbox{for} \; x<0 \\
              \\
            v_R \quad \mbox{for} \; x>0
          \end{array}
           \right.      \label{ri2}
\end{equation}
Ore aim is to solve the RP (\ref{ri2}) within the class of smooth solutions (i. e. in terms of rarefaction waves) so that integration of (\ref{requ}), (\ref{vequ}) along with (\ref{rvin1}) subjected to the initial value problem (\ref{ri2}) will give the required solution. In fact we are going to prove the following 

\vspace{0.2cm}
\noindent
{\bf Theorem 1.} If
\begin{equation}
H\left(  R_{L}^{\gamma}, v_0 (a) \right)=H\left( R_{R}^{\gamma}, v_0 (a) \right) =0 \label{ch}
\end{equation} 
the solution of the RP (\ref{ri2}) is given by
\begin{equation}
R^\alpha \left( \mathbf{U} \right)=R_{L}^{\alpha} \label{soll1}
\end{equation}
and
\begin{eqnarray}
&&v=v_L  \quad \quad \quad \quad \quad \quad  \mbox{if} \quad x < \lambda^N \left( R_{L}^{\gamma}, v_L \right) t  \label{soll2} \\
&& \nonumber \\
&&v=\hat{v} \left( t, v_0 (a) \right), \; x=\int_{0}^{t}{\lambda^N \left( R_L^{\gamma},  \hat{v} \left( t, v_0 (a) \right) \right) dt}, \quad \mbox{if}  \; \lambda^N \left( R_{L}^{\gamma}, v_L \right) t \leq x \leq \lambda^N \left( R_{R}^{\gamma}, v_R \right) t \label{soll3}  \\
&& \nonumber \\
&&  v=v_R  \quad \quad \quad \quad \quad \quad \quad \quad  \mbox{if} \quad  \lambda^N \left( R_{R}^{\gamma}, v_R \right) t < x  \label{soll4}
\end{eqnarray}
provided that
\begin{equation}
R_{L}^{\alpha}=R_{R}^{\alpha}, \quad \quad \quad \lambda^N  \left( R_{L}^{\gamma}, v_L\right) < \lambda^N \left( R_{R}^{\gamma}, v_R \right). \label{j}
\end{equation}
In (\ref{ch}) we set, for simplicity, $H \left( R^\gamma, v \right) = \sigma_{\beta}^{\alpha} \, \mathbf{l}^\beta \cdot \mathbf{B}+ \left( \lambda^N - \lambda^\beta \right)\sigma_{\beta}^{\alpha} \, q^\beta$. Furthermore $\hat{v} \left( t, v_0 (a) \right)$ is given by
\begin{equation}
\frac{d\hat{v}}{dt}=h\left( R_{L}^{\gamma}, \hat{v} \right), \quad \quad \mbox{with} \quad h=B_j + \left( \lambda^N - \lambda^\beta \right)q^\beta d_{j}^{\beta},  \label{tr}
\end{equation}
and $v_0(a)$ denotes the initial datum for $v$ in the singolarity point $(0,0)$.

\vspace{0,2cm}
\begin{proof}
\noindent
Integration of (\ref{requ}), (\ref{vequ}) with the initial data (\ref{ri2}) gives
\begin{eqnarray}
&&R^\alpha=R_{L}^{\alpha}, \quad \quad v=v_L  \quad \quad \quad \quad \mbox{if} \quad x < \lambda^N \left( R_{L}^{\gamma}, v_L \right) t  \label{kk1} \\
&&R^\alpha=R_{R}^{\alpha}, \quad \quad v=v_R  \quad \quad \quad \quad \mbox{if}  
\quad x > \lambda^N \left( R_{R}^{\gamma}, v_R \right) t \label{kk2}
\end{eqnarray} 
Next, in order to connect smoothly the constant left and right states given by (\ref{kk1}), (\ref{kk2}), we solve (\ref{requ}), (\ref{vequ}) with the initial data $R^\alpha (0,0)=R_{0}^{\alpha}(a)$, $v(0,0)=v_0 (a)$ along with the conditions $R_{0}^{\alpha} (0)=R_{L}^{\alpha}$,  $R_{0}^{\alpha}(1)=R_{R}^{\alpha}$, $v_{0}(0)=v_L$, $v_0 (1)=v_R$ with $a \in \left[ 0,1 \right]$. It can be proved (see \cite{cfm} for the proof) that substitution of  the solution of (\ref{requ}), (\ref{vequ}) calculated along the fun of characteristics starting from $(0,0)$ given in (\ref{soll3}) into the constraints (\ref{rvin1}) gives
$$
\frac{dR_{0}^{\alpha}}{da}=0 \quad \quad \Rightarrow \quad R_{0}^{\alpha}=R_{L}^{\alpha}=R_{R}^{\alpha}
$$
and (\ref{soll1}) along with (\ref{j})$_1$ is proved if condition (\ref{ch}) holds. Therefore, since $R^\alpha \left( \mathbf{U} \right) =R_{L}^{\alpha}$, integration of (\ref{vequ}) by the standard method of characteristics produces the solution (\ref{soll2})-(\ref{soll4}) provided that condition (\ref{j})$_2$ is satisfied. 
\end{proof}
\vspace{0.3cm}

\noindent
{\bf Remark 2.} The solution characterizing the rarefaction-like wave will be given in explicit form only if in (\ref{soll3}) it is possible to eliminate $v_0(a)$ . Furthermore, since the behaviour of the rarefaction wave is governed by (\ref{soll3}), such a wave will propagate as in the homogeneous case only when  $h \left( R_{L}^{\alpha}, v_0(a) \right)=0$. We also notice that $R^\alpha \left( \mathbf{U} \right)=R_{L}^{\alpha}$ characterizes the rarefaction curve associated to $\lambda^N$ as it happens for the homogeneous systems so that more general solutions of the RP containing also shock waves and/or contact discontinuities can be obtained as in the homogeneous case. Finally, we notice that in all the cases {\it i)-iii)} of Section $2$, the condition (\ref{ch}) is satisfied so that a Riemann problem can be solved according to Theorem 1.

\vspace{0.2cm}
\noindent
{\bf Remark 3.} If the characteristic speed $\lambda^N$ is exceptional, then it is a Riemann invariant. It follows that $\lambda^N$, calculated in the simple wave region, is constant so that $\lambda^N \left( R_{L}^{\alpha}, v_L\right)=\lambda^N \left( R_{R}^{\alpha}, v_R\right)$. Therefore, integration of (\ref{vequ}) leads to $v=v_L$ if $x<  \lambda^N t$, while $v=v_R$ if $x>\lambda^N t$ and the corresponding solution of (\ref{ri2}) is given by a contact discontinuity.  

\vspace{0.3cm}
As an example, we consider the case of an ideal gas (\ref{rg}) supplemented by the force term (\ref{ccc})$_1$.  Owing to Theorem 1, the solution (\ref{soll1})-(\ref{soll4}) of the concerned RP written in terms of the original variables is
\begin{eqnarray}
&& \rho=\left( \frac{k_0 (\gamma - 1)}{k_2}\left( u - u_L \right) +\rho_{L}^{\frac{\gamma - 1}{2}} \right)^{\frac{2}{\gamma - 1}} \label{sf1} \\
&&u = \left\{
           \begin{array}{l}
            u_L \quad \; \, \quad \; \; \quad \quad \quad \quad \quad  \mbox{if} \quad x< \lambda_{L}^{3} t \\
\\
            \frac{2x}{(\gamma +1)t} +\frac{k_1}{k_0} \quad\; \; \quad \quad \quad \mbox{if} \quad \lambda_{L}^{3}t \leq x \leq \lambda_{R}^{3}t \\
\\
            u=u_R \quad \;  \, \quad \quad \quad \quad \quad \mbox{if} \quad x>\lambda_{R}^{3}t
           \end{array}
           \right. \label{sf2} \\
&&S=S_L \label{sf3}
\end{eqnarray} 
along with the conditions
\begin{eqnarray}
&&u_L=-\frac{k_1}{k_0}+\frac{k_2}{k_0 (\gamma -1)}\rho_{L}^{\frac{\gamma -1 }{2}}, \quad \quad u_R=-\frac{k_1}{k_0}+\frac{k_2}{k_0 (\gamma -1)}\rho_{R}^{\frac{\gamma -1 }{2}}, \quad \quad \rho_L < \rho_R \label{b1} \\
&&S_L=S_R=\hat{S}+C_{V}\ln{\left( \frac{k_2^2}{4 \gamma k_0^2}\right)} \label{b2}
\end{eqnarray}
In (\ref{sf2}) we indicated by $\lambda_L^3 = u_L+\frac{k_2}{2k_0}\rho_{L}^{\frac{\gamma -1}{2}}$ and $\lambda_R^3 = u_R+\frac{k_2}{2k_0}\rho_{R}^{\frac{\gamma -1}{2}}$.  Furthermore, the rarefaction curve associated to $\lambda^3$ is
$$
u=-\frac{k_1}{k_0}+\frac{k_2}{k_0(\gamma - 1)}\rho^{\frac{\gamma - 1}{2}} \quad \quad \mbox{with} \quad \rho_L < \rho.
$$

\vspace{-4cm}
\begin{figure}[h]
\begin{center}
$\begin{array}{c@{\hspace{1in}}c}
\includegraphics[width=3.5in]{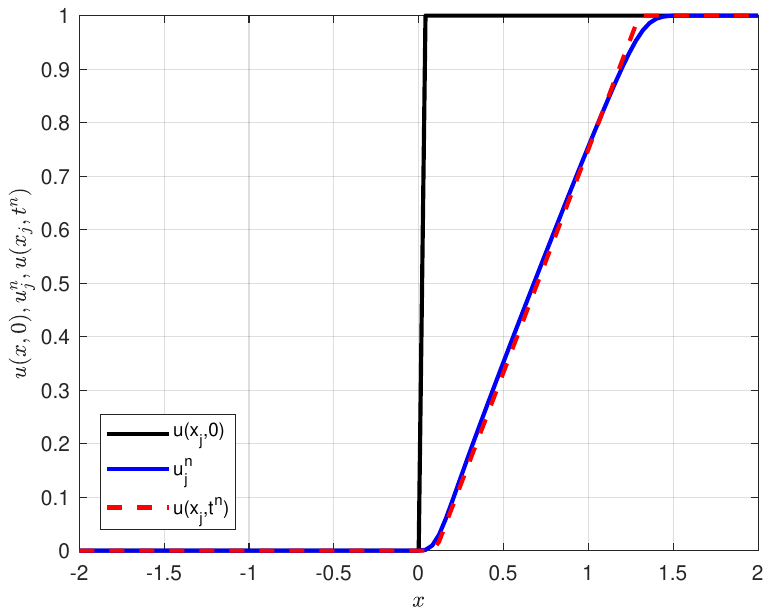} \hspace{-1cm} \includegraphics[width=3.5in]{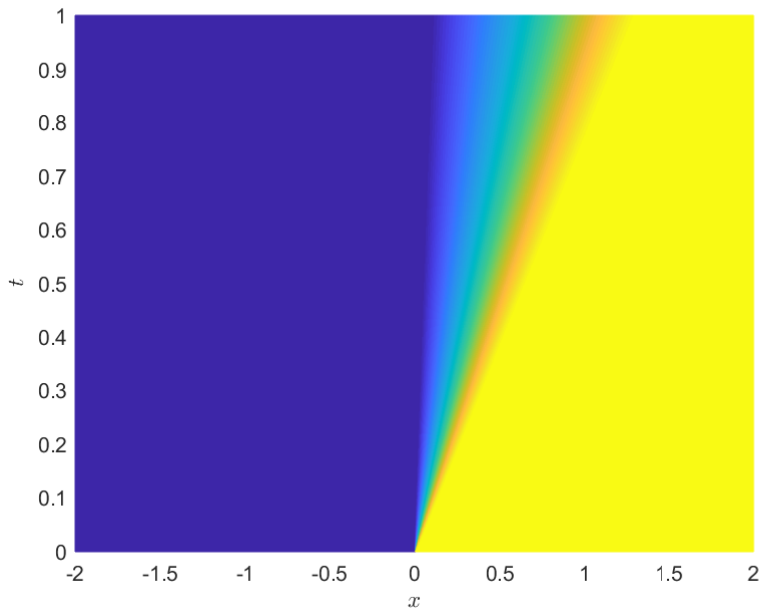}
\end{array}$
\end{center}
\vspace{-4.5cm}
\caption{In the Figure, on the left we give the initial datum, the analytical solution given in (\ref{sf2}) (dotted line) and the numerical solution (solid line) for $u$ for an ideal gas supplemented by (\ref{ccc})$_1$; on the right the rarefaction wave in the $(x,t)$ plane.  }
\end{figure}
  
Finally, to test our procedure, in figure $1$ we show the analytical solution for $u$ given by (\ref{sf2}) and the corresponding numerical solution obtained by integration of (\ref{eu3}).

In order to obtain an accurate numerical solution, without spurious oscillations or excessive numerical dissipation, we propose an High-Resolution Finite Difference Method, with van Leer limiter, constructed using the total variation diminishing property. We integrate the nonlinear equation for $u$ (\ref{eu3}) on a computational domain $[-2,2] \times [0,T]$ at the final time $T=.1$ and step size $\Delta x= 4 \cdot 10^{-2}$ and $\Delta t = 10^{-3}$.

\subsection{Asymptotic behaviour}
Quite recently, Ruggeri and his co-authors\cite{ruggeri1}, \cite{ruggeri2} developed an interesting analysis in the theoretical framework of Riemann problems.   They considered the following hyperbolic system of $N$ PDEs
\begin{eqnarray}
&&\frac{\partial \mathbf{F}^0}{\partial t}+ \frac{\partial \mathbf{F}^i}{\partial x_i}=0  \label{ru1} \\
&&\frac{\partial \mathbf{G}^0}{\partial t}+ \frac{\partial \mathbf{G}^i}{\partial x_i}=\mathbf{f}  \label{ru2}
\end{eqnarray}
where $\mathbf{F}^0 \left( \mathbf{V}, \mathbf{W} \right)$, $\mathbf{F}^i \left( \mathbf{V}, \mathbf{W} \right)$ are column vectors of $\mathbb{R}^M$ ($M<N$), $\mathbf{G}^0 \left( \mathbf{V}, \mathbf{W} \right)$, $\mathbf{G}^i \left( \mathbf{V}, \mathbf{W} \right)$ and $\mathbf{f} \left( \mathbf{V}, \mathbf{W} \right)$ are column vectors of $\mathbb{R}^{N-M}$, while  $\mathbf{V}\in \mathbb{R}^M$ and $\mathbf{W} \in \mathbb{R}^{N-M}$. From the equilibrium states given by $\mathbf{f}\left( \mathbf{V}, \mathbf{W} \right)=0$ we determine $\mathbf{W}=\mathbf{W}\left( \mathbf{V} \right)$ and substituting it into (\ref{ru1}) we obtain the so-called equilibrium sub-system
\begin{equation}
\frac{\partial \mathbf{F}^0 \left( \mathbf{V}, \mathbf{W \left( \mathbf{V} \right)} \right)}{\partial t}+ \frac{\partial \mathbf{F}^i \left( \mathbf{V}, \mathbf{W \left( \mathbf{V} \right)} \right) }{\partial x_i}=0. \label{ru3}
\end{equation}
Furthermore the following sub-characteristics condition must be verified
\begin{equation}
\mbox{min} \; \lambda^k \left( \mathbf{V}, \mathbf{W \left( \mathbf{V} \right)}\right) \leq \mbox{min} \; \mu^\alpha \left( \mathbf{V} \right), \quad \quad  \mbox{max} \; \lambda^k \left( \mathbf{V}, \mathbf{W \left( \mathbf{V} \right)}\right) \geq \mbox{max} \; \mu^\alpha \left( \mathbf{V} \right) \label{sc}
\end{equation}
where $\lambda^k$ (with $k=1,.., N$) are the characteristics velocities of the full system (\ref{ru1}), (\ref{ru2}) evaluated at the equilibrium state while $\mu^\alpha$ (with $\alpha=1,..,M$) are the characteristics velocities of the equilibrium sub-system (\ref{ru3}). 

Thus they conjectured that the solution of a RP for the system (\ref{ru1}), (\ref{ru2}) converges to a combination of shock structures (with or without sub-shocks) and rarefaction waves of the equilibrium sub-system (\ref{ru3}) and they verified such a conjecture numerically.

Here, owing to the results obtained in the previous sections, we aim to give some remarks about the Ruggeri's conjecture. Let us consider the Euler equations (\ref{e1})-(\ref{e3}) where, for generality, we consider the source term $f$ depending also by $S$ . The requirement $f(\rho, u, S)=0$ gives the equilibrium state $u=\hat{u}(\rho , S)$. The resulting equilibrium sub-system is
\begin{eqnarray}
&&\rho_t+\left( \hat{u}+ \rho \hat{u}_\rho \right)\rho_x+\rho \, \hat{u}_S S_x =0 \label{es1} \\
&&S_t+\hat{u} \, S_x =0 \label{es2}
\end{eqnarray}
whose characteristic velocities are $\mu^1 =\hat{u}$ and $\mu^2 =\hat{u} + \rho \hat{u}_\rho$. The Riemann invariant corresponding to $\mu^1$ is $r=\hat{u}(\rho, S)$, while the one corresponding to $\mu^2$ is $r=S$. Since $\mu^1$ is exceptional, we consider the rarefaction wave associated to $\mu^2$ which is characterized by $r=S_0=S_L=S_R$ and it can be calculated  by integration of
\begin{equation}
\rho_t + \left( \hat{u} +\rho \, \hat{u}_\rho \right) \rho_x=0. \label{rare}
\end{equation}
The Riemann invariants of the full system (\ref{e1})-(\ref{e3}) are $R^1=u+\int{\frac{c}{\rho}\, d\rho}$ and $R^2=S$ associated to $\lambda^1=u-c$; $R^1=p(\rho, S)$ and $R^2=u$ corresponding to $\lambda^2=u$;  $R^1=u-\int{\frac{c}{\rho}\, d\rho}$ and $R^2=S$ associated to $\lambda^3=u+c$. Therefore, we notice that one of the  Riemann invariants of the full system, calculated at the equilibrium state, always coincides with the Riemann invariant of the equilibrium sub-system. Furthermore, owing to the analysis developed in the sub-section $4.1$, the rarefaction wave predicted by the full system associated, for instance, to $\lambda^3$ can be determined from
\begin{eqnarray}
&&R^1=u-\int{\frac{c}{\rho}\, d\rho}=R_{0}^{1}=const., \quad \quad R^2=S=R_{0}^{2}=const.  \label{ee1} \\
&&\rho_t + \lambda^3 \rho_x=h\left(R^{1}, R^{2}, u \right).  \label{ee2}
\end{eqnarray}
As far as the sub-characteristics condition (\ref{sc}) is concerned, we notice that (\ref{sc})$_1$ is satisfied while from (\ref{sc})$_2$ we obtain 
\begin{equation}
c \geq \rho\,  \hat{u}_\rho. \label{sb1}
\end{equation}
Therefore, if  $(R^1=R_{0}^{1}, R^2=R_{0}^{2})$ characterizes the equilibrium state and  $h\left( R_{0}^{1}, R_{0}^{2}, u \right)=0 \, \forall u$,  from (\ref{ee1}) we find $R^2=r=S$ and
\begin{equation}
\hat{u}=\int{\frac{c}{\rho}d\rho}+ R_{0}^{1} \label{fi}
\end{equation}
so that  $\lambda^3 \left(\rho, \hat{u}(\rho, S_0), S_0 \right)=\mu^2 \left(\rho, \hat{u}(\rho, S_0) \right)$, the sub-characteristics condition (\ref{sb1}) is satisfied and, in turn, comparing (\ref{rare}) with (\ref{ee2}), we realize that  equation (\ref{ee2}) specializes to (\ref{rare}), the rarefaction wave of the full system propagates as that of the reduced sub-system and the Ruggeri's conjecture is verified. Finally we notice that, owing to (\ref{eu3}), the condition  $h \left( R_{0}^{1}, R_{0}^{2}, u \right)=0$ is verified almost in all the cases  where $f=q^1$ characterized in sub-section $3.1$.

\section{Conclusion and final remarks.}
 Within the theoretical framework of the Method of Differential Constraints, in this paper we develop a reduction procedure for determining exact wave solutions for hyperbolic systems. In particular, by means of a suitable change of variables based on the $k-$Riemann invariants, we simplify the analysis of the compatibility conditions of the overdetermined system composed by the original equations along with the differential constraints. In fact, under suitable structural conditions, in some cases we are able to solve such compatibility equations. The solutions of the governing hyperbolic system generalize the well known simple waves so that they allow to characterize rarefaction-like waves and in turn to solve Riemann problems also for nonhomogeneous models.   

We applied such a procedure  for the Euler equations describing an ideal fluid with a force term depending on the mass density and on the velocity. In the cases of an ideal gas, a Von-K\'arm\'an gas and a Chaplygin gas we characterized classes of force terms  for which wave solutions of the Euler system are determined. The solutions obtained are given in terms of one arbitrary functions so that classes of initial data compatible with the procedure here considered are characterized.

After that, we characterized classes of initial costant states for which the procedure here developed permits to solve Riemann problems by means of generalized rarefaction waves. The behaviour we determine is different from that of the homogeneous case unless the source term $h\left( R^{1}, R^{2}, v \right)$ involved in (\ref{tr}) vanishes when it is calculated in the point $(0,0)$. Furthermore, it is interesting to notice that the corresponding rarefaction curves we obtained are the same of the homogeneous case. Therefore, the general analysis developed for determing the solution of a Riemann problem in terms of rarefaction waves, shock waves and contact discontinuities can be carried on also in the present nonhomogeneous case provided that the system under interest can be written as balance laws.

We end the paper by considering a conjecture made by Ruggeri et al. concerning the asymptotic behaviour of the Riemann problem's solution. In the case of an ideal fluid, owing to the results here obtained, we were able to verify such a conjecture if $h \left( R_{0}^{1}, R_{0}^{2}, v \right)=0$. Of course it remains an open problem to prove such a conjecture for a general system of conservation and balance laws.  On that concerns the use of the $k-$Riemann invariants could be helpful as it was proved in this paper for an ideal gas.

\subsection*{Acknowledgements} N. M. and A. R. thank the financial support of GNFM of the Istituto Nazionale di Alta Matematica. A. J. and N.M. thank also the financial support of University of Messina through project FFABR UNIME 2023. A. J. acknowledges the financial support provided by the project "Strategie HPC e modelli fisico-numerici per la previsione di eventi meteorologici estremi" (HPC-XTREME) from PNRR missione 4 "Istruzione e ricerca" Component 2 "Dalla ricerca all'impresa" - Investment 1.4 - National Center for HPC, Big Data and Quantum Computing (project code CN00000013 - CUP B83C22002830001). A. R. research was partially supported by the PRIN project MIUR Prin 2022, project code 1074 2022M9BKBC, Grant No. CUP B53D23009350006.


\begin{thebibliography}{33}

\bibitem{jan}  Yanenko N N. Compatibility theory and methods of integration of systems of nonlinear partial differential equation. Proc. 4th All-Union Math. Cong. Leningrad: Nauka; 1964, 247-252.

\bibitem{mel1} Meleshko S V.  Methods for constructing exact solutions of partial differential equations. Mathematical and Analytical Techniques with Applications to Engineering. New York: Springer; 2005.

\bibitem{fsy} Fomin V M, Shapeev V P, Yanenko N N. Application of the method of differential constraints to the construction of closed mathematical models, describing one-dimensional dynamic processes in a continuous medium. Chislennye metody mehaniki sploshnoi sredy 1973; 4(3), 39 -47 (Novosibirsk).
	
  \bibitem{sh}  Shapeev V P. Applications of the method of differential constraints to one-dimensional continuum mechanics equation. (PhD thesis) Novosibirsk: Computer center, RAS, 1974.
  
  \bibitem{rsy} Raspopov V E, Shapeev V P, Yanenko N N. Method of differential constraints  for the one-dimensional gas dynamics equations. Chislennye metody mehaniki sploshnoi sredy 1977; 8(2) 100-105 (Novosibirsk).
    
  \bibitem{ms1} Meleshko S V, Shapeev V P. The applications of the differential constraints method to the two-dimensional equations of gas dynamics. J. Appl. Math. Mechs 1999; 63(6), 885-891. https://doi.org/10.1016/S0021-8928(00)00006-X
  
  \bibitem{ms2} Meleshko S V, Shapeev V P. Nonisentropic solutions of simple wave type of the gas dynamics equations. J. Nonlinear Math. Phys. 2011; 18(1) 195-212. https://doi.org/10.1142/S1402925111001374
  
  \bibitem{cfm1} Curr\'o C, Fusco D, Manganaro N. Exact solutions in ideal chromatography via differential constraints method. AAPP – Atti della Accademia Peloritana dei Pericolanti, Classe di Scienze Fisiche, Matematiche e Naturali 2015;  93 (1) A2. https://doi.org/10.1478/AAPP.931A2
  
  \bibitem{cwm} Chaiyasena A,  Worapitpong W, Meleshko S V. Generalized Riemann waves and their adjoinment through a shock wave. Math. Model. Nat. Phenom. 2018;  13(2), 22. https://doi.org/10.1051/mmnp/2018027
  
  \bibitem{cm3} Curr\'o C, Manganaro N. Exact solutions and wave interactions for a viscoelastic medium. AAPP - Atti della Accademia Peloritana dei Pericolanti, Classe di Scienze Fisiche, Matematiche e Naturali 2018; 96 (1) A1. https://doi.org/10.1478/AAPP.961A1
  
  \bibitem{cm4} Curr\'o C,  Manganaro N. Differential constraints and exact solutions for the ET6 model. Ricerche di Matematica 2019; 68, 179-193. https://doi.org/10.1007/s11587-018-0396-6
    
\bibitem{mmw} Meleshko S V, Moyo S, Webb G M.  Solutions of generalized simple wave type of magnetic fluid. Communications in Nonlinear Science and Numerical Simulation 2021;  103, 105991. https://doi.org/10.1016/j.cnsns.2021.105991

\bibitem{mr} Manganaro N, Rizzo A.  Riemann Problems and Exact Solutions for the p-System. Mathematics 2022; 10(67), 935. https://doi.org/10.3390/math10060935.

\bibitem{jmr} Jannelli A, Manganaro N, Rizzo A.  Riemann problems for the nonhomogeneous Aw-Rascle model, Communications in Nonlinear Science and Numerical Simulation 2023; 118, 107010. https://doi.org/10.1016/j.cnsns.2022.107010

\bibitem{mr1} Manganaro N, Rizzo A. Double wave solutions for a hyperbolic model describing nerve fiber. Ricerche di Matematica 2024; 73 (1), 233-245. https://doi.org/10.1007/s11587-023-00792-y

\bibitem{mrv} Manganaro N, Rizzo A, Vergallo P. Solutions to the wave equation for commuting flows of dispersionless PDEs. International Journal of Non-Linear Mechanics 2024; 159, 0104611. https://doi.org/10.1016/j.ijnonlinmech.2023.104611

\bibitem{rive} Rizzo A, Vergallo P. Quasilinear differential constraints for parabolic systems of Jordan-block type. Studies in Applied Mathematics 2025; 154, 6.https://doi.org/10.1111/sapm.70072

\bibitem{ruggeri1}  Brini F,  Ruggeri T. On the perturbed Riemann Problem in Extended Thermodynamics. Rendiconti del Circolo Matematico di Palermo 2006, suppl. 78, 31-44.

\bibitem{ruggeri2}  Mentrelli A, Ruggeri T. Asymptotic Behavior of Riemann and Riemann with Structure Problems for a $2 \times 2$ Hyperbolic Dissipative System. Rendiconti del Circolo Matematico di Palermo 2006, suppl. 78, 201-226.

\bibitem{smo} Smoller J. Shock Waves and Reaction-Diffusion Equation. A Series of Comprehensive Studies in Mathematics, 258 Springer--Verlag, Berlin 1983.

\bibitem{boillat}  Boillat G. The propagation of waves. Traite Phys. Theor. Phys. Math. 23, 1-42, Gauthier-Villar 1965. English translation from La Propagation des Ondes.

\bibitem{lax}  Lax P D. Hyperbolic systems of conservation laws II. Comm. Pure Appl. Math.,  10, 537-566, 1957. https://doi.org/10.1002/cpa.3160100406

\bibitem{cf} Courant R, Friedrichs K. O. Supersonic flow and shock waves. Springer-Verlag 1976.

\bibitem{vk} Von-K\'arm\'an Th. Compressibility effects in aerodynamics. J. of the Aeronautical Science 1941, (9) 8, 337-356.

\bibitem{kmp} Kamenshchik A, Moschella U, Pasquier V. An alternative to quintessence. Physics Letters B 2001, vol 511, 2, 265-268. https://doi.org/10.1016/S0370-2693(01)00571-8


\bibitem{cfm} Curr\'o C, Fusco D, Manganaro N.  Differential constraints and exact solution to Riemann problems for a traffic flow model. Acta Applicandae Mathematicae 2012. Vol. 122, 1, 167-178. https://doi.org/10.1007/s10440-012-9735-x



\end{thebibliography}
\end{document}